\documentclass{article}
\usepackage{amssymb}

\newtheorem{theorem}{Theorem}
\newtheorem{acknowledgement}[theorem]{Acknowledgement}

\input{tcilatex}
\begin{document}

\title{Constructing multi-player quantum games from non-factorizable joint
probabilities}
\author{Azhar Iqbal$^{\text{a,b}}$ and Taksu Cheon$^{\text{a}}$ \\
$^{\text{a}}${\small Kochi University of Technology, Tosa Yamada, Kochi
782-8502, Japan.}\\
$^{\text{b}}${\small Centre for Advanced Mathematics and Physics,}\\
{\small National University of Sciences \& Technology,}\\
{\small campus of College of Electrical \& Mechanical Engineering,}\\
{\small Peshawar Road, Rawalpindi, Pakistan.}}
\maketitle

\begin{abstract}
We use the standard three-party Einstein-Podolsky-Rosen (EPR) setting in
order to play general three-player non-cooperative symmetric games. We
analyze how the peculiar non-factorizable joint probabilities that may
emerge in the EPR setting can change outcome of the game. Our setup requires
that the quantum game attains classical interpretation for factorizable
joint probabilities. We analyze the generalized three-player game of
Prisoner's Dilemma (PD) and show that the players can indeed escape from the
classical outcome of the game because of non-factorizable joint
probabilities. This result for three-player PD contrasts strikingly with our
earlier result for two-player PD for which even non-factorizable joint
probabilities are not found to be helpful to escape from the classical
outcome of the game.
\end{abstract}

\section{Introduction}

The usual approach to quantum games \cite{EWL} considers an initial
(entangled!) quantum state on which players perform local actions
(strategies) and the state evolves to the final state. Payoffs are generated
in the last step involving quantum measurement on the final state. This
approach assumes familiarity with the concepts of product or entangled
quantum state(s), expectation values, trace operation, density operators,
and the theory of quantum measurement.

This paper presents a probabilistic approach to quantum games that
constructs true quantum games from probabilities only. Our motivation has
been to present quantum game to the wider audience, especially to those
readers who use elements of game theory but find the concepts of quantum
mechanics rather alien. We use probabilities to construct quantum games
because, after all, Bell inequalities \cite{Peres} can also be understood in
this way, widely believed to express the true quantum behavior.

We extend our probabilistic framework \cite{IqbalCheon} for two-player
quantum games to mutliplayer case, while using non-factorizable joint
probabilities to construct quantum games. Apart from opening quantum games
to the readers outside of the quantum physics, this framework provides a
unifying perspective on both the classical and the quantum games.

\section{Three-player, two-strategy, non-cooperative, symmetric games}

We consider three-player symmetric games for which players' pure strategies
are given as Alice: $S_{1},S_{2}$; Bob: $S_{1}^{\prime },S_{2}^{\prime }$;
Chris: $S_{1}^{\prime \prime },S_{2}^{\prime \prime }$, and players' payoff
relations are

\begin{equation}
\begin{array}{l}
\Pi _{A,B,C}(S_{1},S_{1}^{\prime },S_{1}^{\prime \prime })=\alpha ,\alpha
,\alpha ; \\ 
\Pi _{A,B,C}(S_{2},S_{1}^{\prime },S_{1}^{\prime \prime })=\beta ,\delta
,\delta ; \\ 
\Pi _{A,B,C}(S_{1},S_{2}^{\prime },S_{1}^{\prime \prime })=\delta ,\beta
,\delta ; \\ 
\Pi _{A,B,C}(S_{1},S_{1}^{\prime },S_{2}^{\prime \prime })=\delta ,\delta
,\beta ;%
\end{array}%
\begin{array}{l}
\Pi _{A,B,C}(S_{1},S_{2}^{\prime },S_{2}^{\prime \prime })=\epsilon ,\theta
,\theta ; \\ 
\Pi _{A,B,C}(S_{2},S_{1}^{\prime },S_{2}^{\prime \prime })=\theta ,\epsilon
,\theta ; \\ 
\Pi _{A,B,C}(S_{2},S_{2}^{\prime },S_{1}^{\prime \prime })=\theta ,\theta
,\epsilon ; \\ 
\Pi _{A,B,C}(S_{2},S_{2}^{\prime },S_{2}^{\prime \prime })=\omega ,\omega
,\omega ,%
\end{array}
\label{symmetric 3-player game definition}
\end{equation}%
where the subscripts refer to the players, the three entries in braces on
left side are pure strategies of Alice, Bob, and Chris, respectively, and
the three entries on right are their payoffs. The three-player Prisoners'
Dilemma offers an example of such a game.

\subsection{Three-player Prisoners' Dilemma}

In this game each of the three players Alice, Bob, and Chris has two pure
strategies: $C$ (Cooperation) and $D$ (Defection). Using the notation
introduced in (\ref{symmetric 3-player game definition}) we associate

\begin{equation}
\text{Alice: }S_{1}\sim C,\text{ }S_{2}\sim D\text{; Bob: }S_{1}^{\prime
}\sim C,\text{ }S_{2}^{\prime }\sim D\text{; Chris: }S_{1}^{\prime \prime
}\sim C,\text{\ }S_{2}^{\prime \prime }\sim D.
\end{equation}%
The three-player Prisoners' Dilemma \cite{3player PD} is defined by
requiring that $S_{2}$ is a dominant choice for each player:

\begin{eqnarray}
\Pi _{A}(S_{2},S_{1}^{\prime },S_{1}^{\prime \prime }) &>&\Pi
_{A}(S_{1},S_{1}^{\prime },S_{1}^{\prime \prime }),  \nonumber \\
\Pi _{A}(S_{2},S_{2}^{\prime },S_{2}^{\prime \prime }) &>&\Pi
_{A}(S_{1},S_{2}^{\prime },S_{2}^{\prime \prime }), \\
\Pi _{A}(S_{2},S_{1}^{\prime },S_{2}^{\prime \prime }) &>&\Pi
_{A}(S_{1},S_{1}^{\prime },S_{2}^{\prime \prime }),  \nonumber
\end{eqnarray}%
and similar inequalities hold for players Bob and Chris. Secondly, a player
is better off if more of his opponents choose to cooperate:

\begin{equation}
\begin{array}{l}
\Pi _{A}(S_{2},S_{1}^{\prime },S_{1}^{\prime \prime })>\Pi
_{A}(S_{2},S_{1}^{\prime },S_{2}^{\prime \prime })>\Pi
_{A}(S_{2},S_{2}^{\prime },S_{2}^{\prime \prime }), \\ 
\Pi _{A}(S_{1},S_{1}^{\prime },S_{1}^{\prime \prime })>\Pi
_{A}(S_{1},S_{1}^{\prime },S_{2}^{\prime \prime })>\Pi
_{A}(S_{1},S_{2}^{\prime },S_{2}^{\prime \prime }).%
\end{array}%
\end{equation}%
Thirdly, if one player's choice is fixed, the other two players are left in
the situation of a two-player PD:

\begin{equation}
\begin{array}{l}
\Pi _{A}(S_{1},S_{1}^{\prime },S_{2}^{\prime \prime })>\Pi
_{A}(S_{2},S_{2}^{\prime },S_{2}^{\prime \prime }), \\ 
\Pi _{A}(S_{1},S_{1}^{\prime },S_{1}^{\prime \prime })>\Pi
_{A}(S_{2},S_{1}^{\prime },S_{2}^{\prime \prime }), \\ 
\Pi _{A}(S_{1},S_{1}^{\prime },S_{2}^{\prime \prime })>(1/2)\left\{ \Pi
_{A}(S_{1},S_{2}^{\prime },S_{2}^{\prime \prime })+\Pi
_{A}(S_{2},S_{1}^{\prime },S_{2}^{\prime \prime })\right\} , \\ 
\Pi _{A}(S_{1},S_{1}^{\prime },S_{1}^{\prime \prime })>(1/2)\left\{ \Pi
_{A}(S_{1},S_{1}^{\prime },S_{2}^{\prime \prime })+\Pi
_{A}(S_{2},S_{1}^{\prime },S_{1}^{\prime \prime })\right\} .%
\end{array}%
\end{equation}%
Using the notation (\ref{symmetric 3-player game definition}) these
conditions require

\begin{equation}
\begin{array}{l}
\text{a) }\beta >\alpha ,\ \ \omega >\epsilon ,\ \ \theta >\delta \\ 
\text{b) }\beta >\theta >\omega ,\ \ \alpha >\delta >\epsilon \\ 
\text{c) }\delta >\omega ,\ \ \alpha >\theta ,\ \ \delta >(1/2)(\epsilon
+\theta ),\ \ \alpha >(1/2)(\delta +\beta )%
\end{array}
\label{Generalized PD Definition}
\end{equation}

\section{Playing three-player games using coins}

The above game can be played using coins and in the following we consider
two setups to achieve this.

\subsection{Three-coin setup}

This setup involves sharing three coins among Alice, Bob, and Chris. We
define pure strategies by the association:

\begin{equation}
\begin{array}{ccc}
S_{1},S_{1}^{^{\prime }},S_{1}^{\prime \prime }\sim \text{flip,} & 
S_{2},S_{2}^{^{\prime }},S_{2}^{\prime \prime }\sim \text{do not flip,} & 
\text{Head}\sim +1,\text{ Tail}\sim -1\text{,}%
\end{array}
\label{association}
\end{equation}%
and play the game as follows. In a run each player receives a coin in `head
up' state which s/he can `flip' or `does not flip'. After players' actions
the coins are passed to a referee. The referee observes the coins and
rewards the players. A player can play a mixed strategy (definable for many
runs) by flipping his/her coin with some probability. This allows us to
write mixed strategies as $(x,y,z)$ where $x$, $y$, and $z$ are the
probabilities with which Alice, Bob, and Chris flip their coins,
respectively. The mixed-strategy payoff relations read

\begin{equation}
\begin{array}{l}
\Pi _{A,B,C}(x,y,z)=xyz(\alpha ,\alpha ,\alpha )+x(1-y)z(\delta ,\beta
,\delta )+xy(1-z)(\delta ,\delta ,\beta )+ \\ 
x(1-y)(1-z)(\epsilon ,\theta ,\theta )+(1-x)yz(\beta ,\delta ,\delta
)+(1-x)(1-y)z(\theta ,\theta ,\epsilon )+ \\ 
(1-x)y(1-z)(\theta ,\epsilon ,\theta )+(1-x)(1-y)(1-z)(\omega ,\omega
,\omega )%
\end{array}%
\end{equation}

Assuming $(x^{\star },y^{\star },z^{\star })$ to be a Nash equilibrium (NE)
requires:

\begin{equation}
\begin{array}{l}
\Pi _{A}(x^{\star },y^{\star },z^{\star })-\Pi _{A}(x,y^{\star },z^{\star
})\geqslant 0, \\ 
\Pi _{B}(x^{\star },y^{\star },z^{\star })-\Pi _{B}(x^{\star },y,z^{\star
})\geqslant 0, \\ 
\Pi _{C}(x^{\star },y^{\star },z^{\star })-\Pi _{B}(x^{\star },y^{\star
},z)\geqslant 0.%
\end{array}
\label{3-coin NE}
\end{equation}

\subsection{Six-coin setup}

This setup translates playing of a three-player game in terms of joint
probabilities which may attain the unusual character of being
non-factorizable for certain quantum systems. The setup thus allows how
non-factorizable (quantum) probabilities lead to game-theoretic consequences.

In this setup, each player receives two coins (either one can be in head or
tail state) and each player chooses one out of the two coins given to
him/her. Players pass three chosen coins to the referee who tosses the three
chosen coins and observes the outcome. After many runs the referee rewards
the players.

Players' strategies are defined by establishing the association:

\begin{equation}
\begin{array}{cc}
S_{1},S_{1}^{^{\prime }},S_{1}^{\prime }\sim \text{choose the first coin,} & 
S_{2},S_{2}^{^{\prime }},S_{2}^{\prime }\sim \text{choose the second coin}%
\end{array}
\label{pure-strategies six-coin setup}
\end{equation}%
A player plays a pure strategy when h/she chooses the same coin for all
runs. He/she plays a mixed-strategy if he/she chooses her/his first coin
with some probability over many runs. We define $x,$ $y,$ and $z$ to be the
probabilities of choosing the first coin by Alice, Bob, and Chris,
respectively.

Note that the quantities $x,$ $y,$ and $z$, though being mathematically
similar in three- and six-coin setups, are physically different in the
following sense. In three-coin setup does not require many runs for the
pure-strategy game. Whereas in the six-coin setup many runs are required for
both the `pure strategy' and the `mixed strategy' games.

In six-coin setup one can define the individual coin probabilities as

\begin{equation}
\begin{array}{c}
r=\Pr (+1;S_{1}),\text{ }r^{\prime }=\Pr (+1;S_{1}^{\prime }),\text{ }%
r^{\prime \prime }=\Pr (+1;S_{1}^{\prime \prime }), \\ 
s=\Pr (+1;S_{2}),\text{ }s^{\prime }=\Pr (+1;S_{2}^{\prime }),\text{ }%
s^{\prime \prime }=\Pr (+1;S_{2}^{\prime \prime }),%
\end{array}%
\end{equation}%
then factorizability of joint probabilities is expressed, for example, as

\begin{equation}
\Pr (+1,-1,-1;S_{2},S_{1}^{\prime },S_{2}^{\prime \prime })=s(1-r^{\prime
})(1-s^{\prime \prime }).
\end{equation}%
We define pure-strategy payoffs by the expressions like

\begin{equation}
\begin{array}{l}
\Pi _{A,B,C}(S_{2},S_{1}^{\prime },S_{1}^{\prime \prime })=(\alpha ,\alpha
,\alpha )sr^{\prime }r^{\prime \prime }+(\delta ,\beta ,\delta
)s(1-r^{\prime })r^{\prime \prime }+(\delta ,\delta ,\beta )sr^{\prime
}(1-r^{\prime \prime })+ \\ 
(\epsilon ,\theta ,\theta )s(1-r^{\prime })(1-r^{\prime \prime })+(\beta
,\delta ,\delta )(1-s)r^{\prime }r^{\prime \prime }+(\theta ,\theta
,\epsilon )(1-s)(1-r^{\prime })r^{\prime \prime }+ \\ 
(\theta ,\epsilon ,\theta )(1-s)r^{\prime }(1-r^{\prime \prime })+(\omega
,\omega ,\omega )(1-s)(1-r^{\prime })(1-r^{\prime \prime })\text{,}%
\end{array}
\label{6coin-payoffs}
\end{equation}%
while the mixed-strategy payoffs are

\begin{equation}
\begin{array}{l}
\Pi _{A,B,C}(x,y,z)=xyz\Pi _{A,B,C}(S_{1},S_{1}^{\prime },S_{1}^{\prime
\prime })+x(1-y)z\Pi _{A,B,C}(S_{1},S_{2}^{\prime },S_{1}^{\prime \prime })+
\\ 
xy(1-z)\Pi _{A,B,C}(S_{1},S_{1}^{\prime },S_{2}^{\prime \prime
})+x(1-y)(1-z)\Pi _{A,B,C}(S_{1},S_{2}^{\prime },S_{2}^{\prime \prime })+ \\ 
(1-x)yz\Pi _{A,B,C}(S_{2},S_{1}^{\prime },S_{1}^{\prime \prime
})+(1-x)(1-y)z\Pi _{A,B,C}(S_{2},S_{2}^{\prime },S_{1}^{\prime \prime })+ \\ 
(1-x)y(1-z)\Pi _{A,B,C}(S_{2},S_{1}^{\prime },S_{2}^{\prime \prime
})+(1-x)(1-y)(1-z)\Pi _{A,B,C}(S_{2},S_{2}^{\prime },S_{2}^{\prime \prime }).%
\end{array}
\label{6-coin mixed-strategy payoffs}
\end{equation}%
A triple $(x^{\star },y^{\star },z^{\star })$ is a NE when%
\begin{equation}
\begin{array}{l}
\Pi _{A}(x^{\star },y^{\star },z^{\star })-\Pi _{A}(x,y^{\star },z^{\star
})\geqslant 0, \\ 
\Pi _{B}(x^{\star },y^{\star },z^{\star })-\Pi _{B}(x^{\star },y,z^{\star
})\geqslant 0, \\ 
\Pi _{C}(x^{\star },y^{\star },z^{\star })-\Pi _{B}(x^{\star },y^{\star
},z)\geqslant 0.%
\end{array}
\label{6-coin NE}
\end{equation}

\subsection{Playing the Prisoner's Dilemma}

Consider playing three-player Prisoner's Dilemma (PD) using the three-coin
setup. The triple $(x^{\star },y^{\star },z^{\star })=(0,0,0)\thicksim
(D,D,D)$ comes out as the unique NE at which the three players are rewarded
as $\Pi _{A}(0,0,0)=\Pi _{B}(0,0,0)=\Pi _{C}(0,0,0)=\omega $. In six-coin
setup we analyze this game when $(s,s^{\prime },s^{\prime \prime })=(0,0,0)$%
, saying that the probability of getting head from each player's second coin
is zero. This reduces the Nash inequalities (\ref{6-coin NE}) to

\begin{equation}
\begin{array}{l}
(x^{\star }-x)\left\{ y^{\star }z^{\star }(rr^{\prime }r^{\prime \prime
})\Delta _{1}+r(z^{\star }r^{\prime \prime }+y^{\star }r^{\prime })\Delta
_{2}+r\Delta _{3}\right\} \geq 0, \\ 
(y^{\star }-y)\left\{ x^{\star }z^{\star }(rr^{\prime }r^{\prime \prime
})\Delta _{1}+r^{\prime }(z^{\star }r^{\prime \prime }+x^{\star }r)\Delta
_{2}+r^{\prime }\Delta _{3}\right\} \geq 0, \\ 
(z^{\star }-z)\left\{ x^{\star }y^{\star }(rr^{\prime }r^{\prime \prime
})\Delta _{1}+r^{\prime \prime }(y^{\star }r^{\prime }+x^{\star }r)\Delta
_{2}+r^{\prime \prime }\Delta _{3}\right\} \geq 0,%
\end{array}
\label{NE-Six-Coin PD}
\end{equation}%
where $\Delta _{1}=(\alpha -\beta -2\delta +2\theta +\epsilon -\omega )$, $%
\Delta _{2}=(\delta -\epsilon -\theta +\omega )$, and $\Delta _{3}=(\epsilon
-\omega ).$ Now for PD we have $\Delta _{3}<0$ and $(D,D,D)$ comes out the
unique NE. This is described by saying that when $(s,s^{\prime },s^{\prime
\prime })=(0,0,0)$ and the joint probabilities are factorizable the triple $%
(D,D,D)$ comes out as the unique NE.

We notice that the requirement $(s,s^{\prime },s^{\prime \prime })=(0,0,0)$
can also be translated as constraints on the joint probabilities involved in
the six-coin setup. For this we first denote these joint probabilities as

\begin{equation}
\begin{array}{l}
p_{1}=rr^{\prime }r^{\prime \prime }\text{,} \\ 
p_{2}=r(1-r^{\prime })r^{\prime \prime }\text{,} \\ 
p_{3}=rr^{\prime }(1-r^{\prime \prime })\text{,} \\ 
p_{4}=r(1-r^{\prime })(1-r^{\prime \prime })\text{,}%
\end{array}%
\text{...}%
\begin{array}{l}
p_{33}=rs^{\prime }s^{\prime \prime }\text{,} \\ 
p_{34}=r(1-s^{\prime })s^{\prime \prime }\text{,} \\ 
p_{35}=rs^{\prime }(1-s^{\prime \prime })\text{,} \\ 
p_{36}=r(1-s^{\prime })(1-s^{\prime \prime })\text{,}%
\end{array}%
\text{...}%
\begin{array}{l}
p_{61}=(1-s)s^{\prime }s^{\prime \prime }\text{,} \\ 
p_{62}=(1-s)(1-s^{\prime })s^{\prime \prime }\text{,} \\ 
p_{63}=(1-s)s^{\prime }(1-s^{\prime \prime })\text{,} \\ 
p_{64}=(1-s)(1-s^{\prime })(1-s^{\prime \prime })\text{,}%
\end{array}
\label{6Coin probs}
\end{equation}%
which allows us to re-express the payoff relations (\ref{6coin-payoffs}) as

\begin{equation}
\begin{array}{l}
\Pi _{A,B,C}(S_{2},S_{1}^{\prime },S_{1}^{\prime \prime })=(\alpha ,\alpha
,\alpha )p_{9}+(\delta ,\beta ,\delta )p_{10}+(\delta ,\delta ,\beta )p_{11}+
\\ 
(\epsilon ,\theta ,\theta )p_{12}+(\beta ,\delta ,\delta )p_{13}+(\theta
,\theta ,\epsilon )p_{14}+(\theta ,\epsilon ,\theta )p_{15}+(\omega ,\omega
,\omega )p_{16}\text{ etc.}%
\end{array}
\label{6coin-1A}
\end{equation}%
Now, in the six-coin setup, the requirement $(s,s^{\prime },s^{\prime \prime
})=(0,0,0)$ makes thirty-seven joint probabilities to vanish:

\begin{equation}
\begin{array}{l}
p_{(9,10,11,12,17,19,21,23,25,26,29,30,33,34,35,37,38,39,41,42,43,44,45,46,49,50,51,52,53,55,57,58,59,60,61,62,63)}=0,%
\end{array}
\label{Constraints on joint probs}
\end{equation}%
which simplifies the pure-strategy payoff relations (\ref{6coin-1A}) to

\begin{equation}
\begin{array}{l}
\Pi _{A,B,C}(S_{1},S_{1}^{\prime },S_{1}^{\prime \prime })=(\alpha ,\alpha
,\alpha )p_{1}+(\delta ,\beta ,\delta )p_{2}+(\delta ,\delta ,\beta )p_{3}+
\\ 
(\epsilon ,\theta ,\theta )p_{4}+(\beta ,\delta ,\delta )p_{5}+(\theta
,\theta ,\epsilon )p_{6}+(\theta ,\epsilon ,\theta )p_{7}+(\omega ,\omega
,\omega )p_{8}; \\ 
\Pi _{A,B,C}(S_{2},S_{1}^{\prime },S_{1}^{\prime \prime })=(\beta ,\delta
,\delta )p_{13}+(\theta ,\theta ,\epsilon )p_{14}+(\theta ,\epsilon ,\theta
)p_{15}+(\omega ,\omega ,\omega )p_{16}; \\ 
\Pi _{A,B,C}(S_{1},S_{2}^{\prime },S_{1}^{\prime \prime })=(\delta ,\beta
,\delta )p_{18}+(\epsilon ,\theta ,\theta )p_{20}+(\theta ,\theta ,\epsilon
)p_{22}+(\omega ,\omega ,\omega )p_{24}; \\ 
\Pi _{A,B,C}(S_{1},S_{1}^{\prime },S_{2}^{\prime \prime })=(\delta ,\delta
,\beta )p_{27}+(\epsilon ,\theta ,\theta )p_{28}+(\theta ,\epsilon ,\theta
)p_{31}+(\omega ,\omega ,\omega )p_{32}; \\ 
\Pi _{A,B,C}(S_{1},S_{2}^{\prime },S_{2}^{\prime \prime })=(\epsilon ,\theta
,\theta )p_{36}+(\omega ,\omega ,\omega )p_{40}; \\ 
\Pi _{A,B,C}(S_{2},S_{1}^{\prime },S_{2}^{\prime \prime })=(\theta ,\epsilon
,\theta )p_{47}+(\omega ,\omega ,\omega )p_{48}; \\ 
\Pi _{A,B,C}(S_{2},S_{2}^{\prime },S_{1}^{\prime \prime })=(\theta ,\theta
,\epsilon )p_{54}+(\omega ,\omega ,\omega )p_{56}; \\ 
\Pi _{A,B,C}(S_{2},S_{2}^{\prime },S_{2}^{\prime \prime })=(\omega ,\omega
,\omega )p_{64}.%
\end{array}
\label{6coin-B}
\end{equation}%
These payoff relations ensure that for factorizable joint probabilities the
classical outcome of the game results.

\section{Three-player quantum games}

We consider three-party EPR setting \cite{Peres} to play a three-player
symmetric game such that each player's two directions of measurement are
his/her pure strategies. In analogy with the six-coin setup, this is
achieved by establishing the association:

\begin{equation}
\begin{array}{l}
S_{1},S_{1}^{^{\prime }},S_{1}^{\prime }\sim \text{choose the first
direction,} \\ 
S_{2},S_{2}^{^{\prime }},S_{2}^{\prime }\sim \text{choose the second
direction.}%
\end{array}%
\end{equation}

Now, in a run, each player chooses one direction out of the two and the
referee is informed about players' choices. The referee rotates
Stern-Gerlach type apparatus along the three chosen directions and performs
(quantum) measurement, the outcome of which, along all the three directions,
is either $+1$ or $-1$.

Comparing the three-party EPR setting to the six-coin setup shows that in a
run, choosing between two directions of measurement\ is similar to choosing
between the two coins. The outcome of (quantum) measurement is $+1$ or $-1$
as it is the case with the coins.

We now denote the joint probabilities in the three-party EPR setting as

\begin{equation}
\begin{array}{l}
p_{1}=\Pr (+1,+1,+1;S_{1},S_{1}^{\prime },S_{1}^{\prime \prime })\text{,} \\ 
p_{2}=\Pr (+1,-1,+1;S_{1},S_{1}^{\prime },S_{1}^{\prime \prime })\text{,} \\ 
p_{3}=\Pr (+1,+1,-1;S_{1},S_{1}^{\prime },S_{1}^{\prime \prime })\text{,} \\ 
p_{4}=\Pr (+1,-1,-1;S_{1},S_{1}^{\prime },S_{1}^{\prime \prime })\text{,}%
\end{array}%
\text{...}%
\begin{array}{l}
p_{61}=\Pr (-1,+1,+1;S_{2},S_{2}^{\prime },S_{2}^{\prime \prime })\text{,}
\\ 
p_{62}=\Pr (-1,-1,+1;S_{2},S_{2}^{\prime },S_{2}^{\prime \prime })\text{,}
\\ 
p_{63}=\Pr (-1,+1,-1;S_{2},S_{2}^{\prime },S_{2}^{\prime \prime })\text{,}
\\ 
p_{64}=\Pr (-1,-1,-1;S_{2},S_{2}^{\prime },S_{2}^{\prime \prime })\text{,}%
\end{array}
\label{EPR Joint Probs}
\end{equation}%
which for coins are reduced to the factorizable joint probabilities involved
in the six-coin setup.

Quantum mechanics imposes constraints on the joint probabilities involved in
the three-party EPR setting. These are usually known as the normalization
and the causal communication constraints \cite{Cereceda}. Normalization says
that

\begin{equation}
\begin{array}{l}
\sum_{i=1}^{8}p_{i}=1,\text{ }\sum_{i=9}^{16}p_{i}=1,\text{...}%
\sum_{i=57}^{64}p_{i}=1\text{.}%
\end{array}
\label{normalization}
\end{equation}%
While, the causal communication constraint is expressed as

\begin{equation}
\begin{array}{c}
\tsum\nolimits_{i=1}^{4}p_{i}=\tsum\nolimits_{i=17}^{20}p_{i}=\tsum%
\nolimits_{i=25}^{28}p_{i}=\tsum\nolimits_{i=33}^{36}p_{i} \\ 
\tsum\nolimits_{i=5}^{8}p_{i}=\tsum\nolimits_{i=21}^{24}p_{i}=\tsum%
\nolimits_{i=29}^{32}p_{i}=\tsum\nolimits_{i=37}^{40}p_{i} \\ 
\tsum\nolimits_{i=9}^{12}p_{i}=\tsum\nolimits_{i=41}^{44}p_{i}=\tsum%
\nolimits_{i=49}^{52}p_{i}=\tsum\nolimits_{i=57}^{60}p_{i} \\ 
\tsum\nolimits_{i=13}^{16}p_{i}=\tsum\nolimits_{i=45}^{48}p_{i}=\tsum%
\nolimits_{i=53}^{56}p_{i}=\tsum\nolimits_{i=61}^{64}p_{i}%
\end{array}
\label{locality-1}
\end{equation}

\begin{equation}
\begin{array}{c}
\tsum\nolimits_{i=1}^{4}p_{2i-1}=\tsum\nolimits_{i=5}^{8}p_{2i-1}=\tsum%
\nolimits_{i=13}^{16}p_{2i-1}=\tsum\nolimits_{i=21}^{24}p_{2i-1} \\ 
\tsum\nolimits_{i=1}^{4}p_{2i}=\tsum\nolimits_{i=5}^{8}p_{2i}=\tsum%
\nolimits_{i=13}^{16}p_{2i}=\tsum\nolimits_{i=21}^{24}p_{2i} \\ 
\tsum\nolimits_{i=9}^{12}p_{2i-1}=\tsum\nolimits_{i=17}^{20}p_{2i-1}=\tsum%
\nolimits_{i=25}^{28}p_{2i-1}=\tsum\nolimits_{i=29}^{32}p_{2i-1} \\ 
\tsum\nolimits_{i=9}^{12}p_{2i}=\tsum\nolimits_{i=17}^{20}p_{2i}=\tsum%
\nolimits_{i=25}^{28}p_{2i}=\tsum\nolimits_{i=29}^{32}p_{2i}%
\end{array}
\label{locality-2}
\end{equation}

\begin{equation}
\begin{array}{l}
p_{1}+p_{2}+p_{5}+p_{6}=p_{17}+p_{18}+p_{21}+p_{22}=p_{9}+p_{10}+p_{13}+p_{14}=p_{49}+p_{50}+p_{53}+p_{54},
\\ 
p_{3}+p_{4}+p_{7}+p_{8}=p_{19}+p_{20}+p_{23}+p_{24}=p_{11}+p_{12}+p_{15}+p_{16}=p_{51}+p_{52}+p_{55}+p_{56},
\\ 
p_{25}+p_{26}+p_{29}+p_{30}=p_{33}+p_{34}+p_{37}+p_{38}=p_{41}+p_{42}+p_{45}+p_{46}=p_{57}+p_{58}+p_{61}+p_{62},
\\ 
p_{27}+p_{28}+p_{31}+p_{32}=p_{35}+p_{36}+p_{39}+p_{40}=p_{43}+p_{44}+p_{47}+p_{48}=p_{59}+p_{60}+p_{63}+p_{64}.%
\end{array}
\label{locality-3}
\end{equation}

Essentially, these constraints state that, in a run, on referee's
measurement, the outcome of $+1$ or $-1$ along Alice's chosen direction is
independent of what choices Bob and Chris make for their directions. The
same applies for Bob and Chris.

Notice that the factorizable joint probabilities also satisfy the causal
communication constraint as do the three-party EPR joint probabilities.
Whereas, unlike the coin probabilities, the three-party EPR joint
probabilities can be non-factorizable. In this case if (\ref{EPR Joint Probs}%
) are expressed as (\ref{6Coin probs}) one of more of the probabilities $r,$ 
$r^{\prime },$ $r^{\prime \prime },$ $s,$ $s^{\prime },$ $s^{\prime \prime }$
becomes negative or greater than one.

\subsection{Three-player quantum Prisoner's Dilemma}

Notice that the constraints (\ref{Constraints on joint probs}) ensure that
for a factorizable joint probabilities the triple $(D,D,D)$ becomes a NE and
that requiring that a set of (quantum mechanical) joint probabilities to
satisfy the constraints (\ref{Constraints on joint probs}) imbeds the
classical game within the corresponding quantum game.

We now consider playing the three-player PD using the three-party EPR
setting. We ask whether the triple $(C,C,C)$ can be a NE for
non-factorizable joint probabilities while our setup ensures that for
factorizable three-party EPR joint probabilities the game can be interpreted
classically, with the triple $(D,D,D)$ being its unique NE. To answer this
we use (\ref{6coin-B}), (\ref{normalization}), and (\ref{locality-1}-\ref%
{locality-3}) to find the NE from (\ref{6-coin NE}) and allow the involved
joint probabilities to become non-factorizable.

For two-player PD we have reported that $(D,D,D)$ once again emerges as the
unique NE with the same definition of players' strategies and under the
requirements that embed the classical game within the quantum.

For three-player PD the situation, however, comes out to be different. The
Nash inequalities for the triple $(C,C,C)$ then read

\begin{equation}
\begin{array}{l}
\left\{ p_{5}+(\alpha /\beta )p_{1}-p_{13}\right\} +(\theta /\beta )\left\{
p_{6}+p_{7}-p_{14}-p_{15}+(\delta /\theta )(p_{2}+p_{3})\right\} + \\ 
(\omega /\beta )\left\{ p_{8}-p_{16}+(\epsilon /\omega )p_{4}\right\} \geq 0;%
\end{array}
\label{(C,C,C)NE(A)}
\end{equation}

\begin{equation}
\begin{array}{l}
\left\{ p_{2}+(\alpha /\beta )p_{1}-p_{18}\right\} +(\theta /\beta )\left\{
p_{4}+p_{6}-p_{20}-p_{22}+(\delta /\theta )(p_{3}+p_{5})\right\} + \\ 
(\omega /\beta )\left\{ p_{8}-p_{24}+(\epsilon /\omega )p_{7}\right\} \geq 0;%
\end{array}
\label{(C,C,C)NE(B)}
\end{equation}

\begin{equation}
\begin{array}{l}
\left\{ p_{3}+(\alpha /\beta )p_{1}-p_{27}\right\} +(\theta /\beta )\left\{
p_{4}+p_{7}-p_{28}-p_{31}+(\delta /\theta )(p_{2}+p_{5})\right\} + \\ 
(\omega /\beta )\left\{ p_{8}-p_{32}+(\epsilon /\omega )p_{6}\right\} \geq 0;%
\end{array}
\label{(C,C,C)NE(C)}
\end{equation}%
where $\alpha /\beta ,$ $\theta /\beta ,$ $\delta /\theta ,$ $\omega /\beta ,
$ $\epsilon /\omega $ $<1$. We find that a set of (quantum) joint
probabilities that satisfy the normalization and the causal communication
constraints can indeed allow the inequalities (\ref{(C,C,C)NE(A)}-\ref%
{(C,C,C)NE(C)}) to be true. For example, take $\alpha /\beta =9/10,$ $\theta
/\beta =1/100,$ $\delta /\theta =1/5,$ $\omega /\beta =1/100,$ $\epsilon
/\omega =9/10$ and assign values to these joint probabilities as $p_{1}=1/10,
$ $p_{3}=13/100,$ $p_{5}=16/100,$ $p_{6}=1/10,$ $p_{13}=14/100,$ $p_{15}=2/5,
$ $p_{18}=13/100,$ $p_{20}=1/4,$ $p_{22}=37/100,$ $p_{27}=1/5$ which we call
as the `independent probabilities'. Notice that constraints (\ref%
{Constraints on joint probs}) assign zero value to thirty seven joint
probabilities out of the remaining ones and using the normalization and
causal communication constraints the values assigned to the rest of joint
probabilities are then found as $p_{2}=7/50,$ $p_{4}=1/100,$ $p_{7}=3/20,$ $%
p_{8}=21/100,$ $p_{14}=9/25,p_{16}=1/10,$ $p_{24}=1/4,$ $p_{28}=9/50,$ $%
p_{31}=17/50,$ $p_{32}=7/25,p_{36}=19/50,$ $p_{40}=31/50,$ $p_{47}=27/50,$ $%
p_{48}=23/50,$ $p_{54}=1/2,$ $p_{56}=1/2$. With this the above NE
inequalities for $(C,C,C)$ reduce to $0.106\geq 0,$ $0.096\geq 0,$ $%
0.017\geq 0$ which are trivially true.

Note that for PD we have $\alpha /\beta ,$ $\theta /\beta ,$ $\delta /\theta
,$ $\omega /\beta ,$ and $\epsilon /\omega $ all less than zero and not
every non-factorizable set of joint probabilities can result in $(C,C,C)$
being a NE. In this paper we do not explore which other NE may emerge for a
given non-factorizable set of probabilities. However, we notice that the
classical outcome of $(D,D,D)$ being a NE remains intact even when the joint
probabilities may become non-factorizable. This means that a set of
non-factorizable joint probabilities can only add to the unique classical NE
in the three-player PD.

\section{Concluding remarks}

We use three-party EPR setting to play a three-player symmetric
noncooperative game. Players' payoffs are re-expressed in terms of players'
choices in the EPR setting and in terms of the joint probabilities. We use
Nash inequalities in the six-coin setup to impose constraints on joint
probabilities which ensure that with factorizable joint probabilities\ the
game has a classical interpretation. We then retain these constraints while
allowing the joint probabilities to become non-factorizable and find how
non-factorizable probabilities may lead to the emergence of new solutions of
the game. Multi-player quantum games are, therefore, constructed in terms of
probabilities only and it is shown that non-factorizable joint probabilities
may lead to different game-theoretic outcome(s). We find that with this
framework it is hard to construct Enk \& Pike type argument \cite{EnkPike}
for a quantum game.

\begin{acknowledgement}
One of us (AI) is supported by the Japan Society for Promotion of Science
(JSPS). This work is supported, in part, by the Grant-in-Aid for scientific
research provided by Japan Ministry of Education, Culture, Sports, Science,
and Technology under the contract numbers 18540384 and 18.06330.
\end{acknowledgement}

\end{document}